\documentclass{egpubl}
\usepackage{eurovis2021}

\SpecialIssuePaper         

\usepackage[T1]{fontenc}
\usepackage{dfadobe}  

\usepackage{cite}  

\usepackage{amsmath}
\usepackage{mathtools}

\usepackage{cite,url,mathptmx,times,amsmath,algorithmic,cases,color,subfigure}

\usepackage{tikz}
\usetikzlibrary{calc}

\usepackage[font=small,labelfont=bf,tableposition=top]{caption}

\usepackage{cuted}
\usepackage{capt-of}
\usepackage{array}
\usepackage{wrapfig}
\usepackage{calc}

\def\_{\rule{.3em}{.15ex}}      

\newtheorem{definition}{Definition}
\newtheorem{property}[definition]{Property}
\newtheorem{proposition}[definition]{Proposition}
\newtheorem{lemma}[definition]{Lemma}
\newtheorem{theorem}[definition]{Theorem}
\newtheorem{corollary}[definition]{Corollary}

\newcommand {\mymarginpar}[1]{\marginpar{#1}}
\renewcommand {\marginpar}[1]{}

\newcommand {\rfig}[1]{Figure \ref{fig:#1}}

\newcommand {\bsec}[2]{\section{#1}
                       \label{sec:#2} }

\newcommand {\rsec}[1]{Section \ref{sec:#1}}

\newcommand {\bsubsec}[2]{\mymarginpar{sec:#2}
                       \subsection{#1}
                       \label{sec:#2} }

\newcommand {\rsubsec}[1]{Section \ref{sec:#1}}

\newcommand {\beq}[1]{
                      \begin{equation}
                      \label{eq:#1} }

\newcommand {\beqno}[1]{\begin{eqnarray}
                      \nonumber}

\newcommand {\eeq}       {\end{equation}}
\newcommand {\eeqno}       { && \end{eqnarray}}
\newcommand {\req}[1]{Eq.~(\ref{eq:#1})}

\newcommand {\bear}[1]{
                       \begin{eqnarray}
                       \label{eq:#1} }

\newcommand {\bearno}[1]{
                       \begin{eqnarray}
                       \nonumber}

\newcommand {\eear}{\end{eqnarray}}
\newcommand {\eearno}{\end{eqnarray}}

\newcommand {\btab}[1]{
                       \begin{table}
                       \centering
                       \begin{tabular}{#1}}
\newcommand {\etab}[3] {
                       \end{tabular}
                       \caption[#3]{#2}
                       \label{tab:#1}
                       \end{table}
                       \vspace{.1in}}

\newcommand {\btabular}[1]{\begin{center}
                       \begin{tabular}{#1}}
\newcommand {\etabular}{\end{tabular}
                       \end{center}}

\newcommand {\bdefin}[1]{\begin{definition}\label{def:#1}}
\newcommand {\edefin}       {\end{definition}}

\newcommand {\bpro}[1]{\begin{property}
                      \label{pro:#1} }
\newcommand {\epro}   {\end{property}}

\newcommand {\bprop}[1]{\begin{proposition}
                      \label{prop:#1} }
\newcommand {\eprop}       {\end{proposition}}

\newcommand {\blem}[1]{\begin{lemma}
                      \label{lem:#1}}
\newcommand {\elem}   {\end{lemma}}

\newcommand {\bthe}[1]{\begin{theorem}
                      \label{the:#1} }
\newcommand {\ethe}   {\end{theorem}}

\newcommand {\bcor}[1]{\begin{corollary}
                      \label{cor:#1} }
\newcommand {\ecor}   {\end{corollary}}

\newcommand{\hide}[1]{}

\newcommand{\shil}[1]{{\color{red}[From Lei: #1]}}

\newcommand{\codename}{iQUANT }
\newcommand{\codenamenospace}{iQUANT}

\newcommand{\eurovis}[1]{\textcolor{black}{#1}}
\newcommand{\rebuttal}[1]{\textcolor{black}{#1}}
\BibtexOrBiblatex
\electronicVersion
\PrintedOrElectronic

\usepackage{egweblnk}

\title[iQUANT: Interactive Quantitative Investment Using Sparse Regression Factors]%
      {iQUANT: Interactive Quantitative Investment Using Sparse Regression Factors}

\author[Xuanwu Yue et al.]
{\parbox{\textwidth}{
    \centering Xuanwu Yue$^{1,4}$\orcid{0000-0002-9714-6545},
        Qiao Gu$^{2}$\orcid{0000-0002-1342-3307},
        Deyun Wang$^{3}$\orcid{0000-0001-9817-361X},
        Huamin Qu$^{4}$\orcid{0000-0002-3344-9694}
        and Yong Wang$^{5}$\orcid{0000-0002-0092-0793}
        }
        \\
{\parbox{\textwidth}{\centering 
        $^1$Sinovation Ventures AI Institute, China \\
        $^2$Robotics Institute, Carnegie Mellon University, USA \\
        $^3$Institute of Software, Chinese Academy of Sciences, China \\
        $^4$The Hong Kong University of Science and Technology, Hong Kong, China \\
        $^5$School of Computing and Information Systems, Singapore Management University, Singapore
      }
}
}
\begin{document}


\maketitle
\begin{abstract}
The model-based investing using financial factors is evolving as a principal method for quantitative investment. The main challenge lies in the selection of effective factors towards excess market returns. Existing approaches, either hand-picking factors or applying feature selection algorithms, do not orchestrate both human knowledge and computational power. This paper presents iQUANT, an interactive quantitative investment system that assists equity traders to quickly spot promising financial factors from initial recommendations suggested by algorithmic models, and conduct a joint refinement of factors and stocks for investment portfolio composition. We work closely with professional traders to assemble empirical characteristics of ``good'' factors and propose effective visualization designs to illustrate the collective performance of financial factors, stock portfolios, and their interactions. We evaluate iQUANT through a formal user study, two case studies, and expert interviews, using a \eurovis{real} stock market dataset consisting of 3000 stocks $\times$ 6000 days $\times$ 56 factors.



\end{abstract}  

\bsec{Introduction}{Intro}


Quantitative investment using algorithmic trading has accounted for over 85\% of transactions in US stock markets since 2012 \cite{glantz2013multi}. There is a growing demand for machine learning models that can describe and explain the price change of an individual or a portfolio of stocks in the market.
\eurovis{Factor investing is one of the most widely used strategies in quantitative investment.}
In a factor investing paradigm \cite{TsinghuaSurvey, ThreeFactor, CAPM}, a selective collection of financial variables called \textit{factors} (e.g., a company's fundamental data and its growth measures) are used as input predictors and stock returns are set as target outcomes to build predictive models. With these models, traders forecast future stock returns regularly in order to construct time-sensitive portfolios composed of a small set of high-return securities. The returns of these securities are expected to beat the average market return, which is known as the excess return. Recent global surveys show that the factor investing paradigm is employed by 90\% of institutional investors to manage at least part of their portfolio \cite{FactorRising}. The percentage of assets traded with factor investing is over 30\% in size and still fast-growing.



Despite the prevalence of factor investing, selecting an effective set of factors for market investment is a difficult task. There can be hundreds of candidate factors or more available for predictive models. Currently, most quantitative traders
manually select factors by comparing their corresponding portfolio returns in a standard backtesting.
Such a trial-and-error approach is indirect, less intuitive, and often requires an exhaustive search in the vast combinatorial factor space. To compute returns from a set of factors, a predictive model and stock portfolio need to be constructed, whose interactions with the selected factors are not fully understood and utilized in the current approach. Traders must also speculate an initial factor selection for evaluation. On the other hand, feature selection methods in machine learning research \cite{guyon2003introduction} have been applied to this domain, but not as an end-to-end approach like many other fields. Algorithm-selected factors are mostly used as a reference or starting point of the manual selection process, as these factors (especially in black-box models) are not trusted by traders until they can be explained and evaluated. There \rebuttal{are} currently few tools that can combine human intelligence (e.g., prior knowledge of a factor's utility) and computational power (e.g., assessing stability and sensitivity of a factor) for factor selection in stock markets. 

\vspace{-0.1in}
This work is committed to three essential requirements of interactive factor investing. First, traders need an intuitive and comprehensive way to assess the performance of numerous factors with respect to excess returns. Second, high-performance factors need to be selected and refined together with high-return stocks in the construction of an investment portfolio. Third, portfolios need to be evaluated and compared in the industry-standard backtesting for their returns in both past and future outlook. To meet these requirements, we present \codename (\underline{i}nteractive \underline{Q}uantitative investment \underline{U}sing sp\underline{A}rse regressio\underline{N} fac\underline{T}ors), a visual analytics system that seamlessly integrates human and algorithmic factor selection for effective factor investing in stock markets. \codename enjoys the best of both worlds from human intelligence by a heuristic search in the combinatorial factor space and from computational power by summarizing and presenting key performance metrics of factors and stocks in close interactions. The contributions of this work can be summarized as follows.



\begin{itemize}
 \item We explore the use of sparse regression models as the initial feature selection method for factor investing. Sparse regression models are preferred as they optimize the procedure between factor selection and stock prediction in the same process. Performance metrics of factors are computed, including not only their feature weights, but also the factor contribution that describes its importance to the prediction.

 \item We introduce elaborate visualization designs that comprehensively illustrate the performance of factors in the regression model to predict the prices of candidate stocks. Our designs are based on expert studies about empirical characteristics of valuable factors, including stability and  sensitivity.
 
 \item On top of the visualization interface, we propose an interaction framework for factor investing that allows traders to jointly select high-performance factors and high-return stocks in an integrated model-building and portfolio-construction process. By our framework, traders can iteratively refine the factor selection and stock portfolio based on visual evidence shown in the interface. The industry-standard backtesting is also supported.

\item We evaluate \codename in a formal user study comparing with baseline alternatives. 
Case studies with domain experts on targeted user tasks are conducted \eurovis{using a real stock market dataset}.
Expert feedback about 
\codename is also reported and discussed. 

\end{itemize} 
\vspace{-0.05in}
\bsec{Related Work}{Related}

\bsubsec{Financial Data Visualization}{Financial-Vis}
Financial data, such as stock/fund prices and economic indicators, has been widely used by 
traders and investors to guide their investment decisions. Many visualization approaches have been proposed to support interactive analysis of financial data in the past decades~\cite{EuroVis16Survey,FinanceVis.net,BusinessVisSurvey,rodriguez2016visualizing,tsang2020tradao,lin2020taxthemis}. 
Since the target users are often not visualization experts, classical designs such as line charts and bar charts are popular in financial visualization for better usability~\cite{WireVis,ziegler2010visual,BloombergVis,BitExTract,sportfolio}. 
For example, Chang et al.~\cite{WireVis} applied multiple coordinated views with enhanced line charts and matrix designs to illustrate the time-varying wire transactions in banking.
Ziegler et al.~\cite{ziegler2010visual} designed color-coded bars to display relative changes in asset prices over time. Sorenson and Brath~\cite{BloombergVis} combined continuous line charts with icon-based event representations to inform users about the important context of financial data. With the ever-increasing data size, financial visualization methods are often integrated with automatic analytics to support visual analysis of financial data~\cite{vsimunic2003visualization,schreck2007trajectory,schreck2009visual,van1999cluster,ziegler2010visual}. Meanwhile, versatile visualizations beyond classical charts have also been developed to cope with the complexity of modern financial data, e.g., Growth Matrix \cite{keim2006spectral,ziegler2007relevance}, wedge charts \cite{dao2008nasdaq}, and 3D visualizations \cite{tekusova2008visualizing,nesbitt2004finding}.

Currently, few studies have been done in integrating visualization into the factor investing process. 
Some existing studies~\cite{FundExplorer,FinVis,savikhin2011experimental} apply visualizations to help manage investment portfolios. FundExplorer~\cite{FundExplorer} introduces a distorted treemap to visualize both the composition of a fund's portfolio and the remaining stocks. Their design facilitates the convenient retrieval of complementary funds to achieve portfolio diversification.
FinVis~\cite{FinVis} employs visualization to help non-experts understand the expected returns, risks and the changing context of multiple portfolios.
PortfolioCompare~\cite{savikhin2011experimental} displays the variability and correlation of a portfolio's risks and returns by scatterplots and distribution charts. 
Their designs can assist users in making better investments according to their risk preferences. None of these visualization 
tools is integrated with predictive and feature selection models for factor investing. 

\vspace{-0.15in}
\bsubsec{Interactive Feature Selection}{Rel-Selection}
Feature selection is a key step in machine learning. Prior studies have focused on algorithmic feature selections~\cite{li2018featureSelectionSurvey}, whose mechanism and reasoning are often opaque to users.
Inspired by prior research on 
predictive model visualization~\cite{liu2018visual,puri2020rankbooster,zhao2019iforest,ming2019rulematrix,el2019visual},
interactive feature selections have been widely studied. These studies allow users to visually explore the feature space by novel representations and interactive interfaces. Seo and Shneiderman~\cite{seo2005rankByFeature} proposed a rank-by-feature framework for the exploration of multidimensional data. Data features are ranked and visualized by scatter plots, bar charts, etc. Johansson and Johansson~\cite{johansson2009interactive} utilized parallel coordinates to visualize data features, with the order determined by user-defined metrics.

On the other hand, there are some works considering the relationship between features and predictive models. For example, SmartStripes~\cite{May2011smartStripes} combines interactive feature selection by allowing users to explore the dependencies between features and algorithms. INFUSE~\cite{krause2014infuse} integrates feature selection with prediction models and a glyph-based design to inform users about the feature performance with respect to the prediction performance. iFEED~\cite{bang2016ifeed} and FeatureMiner~\cite{cheng2016featureminer} visualize the feature performance by line charts, bar charts and scatter plots. Prospector~\cite{krause2016interacting} applies partial dependence diagnostics to analyze the correlation among features, data values, and prediction results. RegressionExplorer \cite{RegressionExplorer} combines feature space visualization of regression models with a dynamic subgroup of data analysis.

Existing studies summarize the feature dependency or performance by hybrid quality measures (e.g., feature ranking), but the temporal evolution of features are rarely considered. In factor investing, the temporal dynamics of financial factors (features) and their influence on the prediction model are crucial, which makes existing work not suitable for our scenario.

\vspace{-0.1 in}
\bsec{System Overview}{System}

In this section, the technical background of interactive factor investing is introduced, on which key user requirements are summarized. To meet these requirements, four user tasks are defined which are achieved by a pipeline design in our system.


\vspace{-0.1 in}
\bsubsec{Background}{Scenario}


Historically, factor investing originated from the Capital Asset Pricing Model (CAPM) proposed by \rebuttal{Sharpe \cite{Sharpe1964CAPM} and Linter\cite{linter1965CAPM}}, which uses a single factor of a stock's sensitivity to market returns to explain stock returns. Then, researchers and practitioners continued to discover new factors related to stock returns. Notably, Fama and French~\cite{ThreeFactor} proposed the famous three-factor model which considers a size factor (large v.s. small capitalization stocks) and a value factor (high v.s. low book-to-market ratios) in addition to the market sensitivity factor. Later, many factors explaining stock returns were discovered, including a company's recent fundamental data \cite{asness1997interaction}, the long-term income growth of companies \cite{porta1998law, dechow1997returns}, etc.

In a typical quantitative investment scenario for stock markets, traders hold a portfolio composed of a collection of individual stocks appropriately mixed in value. The portfolio is adjusted in regular time intervals, which is also known as the trading cycle. Depending on the style of traders and funds, the trading cycle can range from one day/week (e.g., private equity) to several months (e.g., public funds). By the factor investing paradigm, at the beginning of each trading cycle, traders pick a few factors based on their experience, domain knowledge, market status and factor returns. 
Quantitative models are then built up over historical market data, which try to establish linkages between factors and stock returns. Both machine learning and statistical models have been applied in this stage\rebuttal{\cite{coqueret2020machine}}. From model-predicted stock returns, traders further apply a trading strategy to determine how to adjust their investment assets, i.e., when and which stocks to buy/sell. Because it is extremely hard to forecast how the portfolio adjustment decision performs in the upcoming trading cycle, a standard backtesting helps to evaluate its returns on historical data. The model consistently earns excess returns to the market, or at least in recent cycles, is considered good in the context of quantitative investment. 


\vspace{-0,1in}
\bsubsec{Task Characterization}{Task}

The factor investing paradigm introduced above seems straightforward. However, due to the closedness of the quantitative investment community, we only came to this understanding after pilot studies with expert traders in several mutual funds, both public and private ones. In the interviews, almost all of them mentioned a key difficulty in the factor investing practice, the selection of valuable factors. While during their long-term jobs, a number of factor combinations proven in the past have been accumulated, they never stop finding and evaluating new factors because stock fluctuations in the future can hardly be entirely the same as the past. In a sense, the ability to discover new factors is the central competence of quantitative traders and accounts for most of their performance difference.

As we learned from experts in the interview, the current factor selection process is mostly done manually. A major user pain point lies in that the evaluation of selected factors are indirect, less intuitive, and often requires an exhaustive search in the vast combinatorial factor space to identify the best factor collection. In the standard backtesting of quantitative models, to be able to compute model returns, the learning/statistical algorithm and the focused stock need to be determined in addition to the factor selection. Importantly, these model returns are often the result of interactions among factor, model and stock selections. Users often require a tedious trial-and-error comparison process to determine factors. As such a process is done in every trading cycle, efficiency is also a critical consideration not to be neglected. Meanwhile, automatic factor selections by machine learning algorithms (e.g., feature selection) have been incorporated in a trader's job. Due to explainability concerns, these algorithm-picked factors are only used as a starting point or reference for manual factor selection until traders can understand their working mechanism. To this end, interactive factor investing systems that can integrate feature selection, model building, and human intelligence in heuristic factor search, as well as visually explain selected factors are of great demand. Below we summarize key user requirements collected from our expert studies.

\textbf{R1. Direct and intuitive factor evaluation.} As there are a large number of candidate factors, for a given pool of stocks, the system should recommend an initial factor selection for users to start with (\textbf{R1.1}). Importance measures of these factors should be computed in the context of quantitative models to account for the interaction between factors and models. These factor importance and the performance of underlying stock/model need to be presented comprehensively to allow intuitive factor evaluation (\textbf{R1.2}).

\textbf{R2. Interaction support for factor and portfolio selection.} Based on evaluations of factor and stock performance, the system should support appropriate interaction to help construct stock portfolios to invest. This requires interactions to jointly select and refine factors and stocks, as well as a mechanism to help users understand.

\textbf{R3. Integrated model building and backtesting.} In the factor selection and evaluation, the system should automatically train quantitative models to predict stock returns based on selected factors and stocks, algorithmically or jointly (\textbf{R3.1}). In factor investing, backtesting of these models should be performed to compare portfolio returns and guide investment decisions (\textbf{R3.2}).

To meet the requirements of expert traders, we target to build a visual analytics system that can support the following key tasks.

\textbf{T1. Selection of an initial pool of stocks and factors.} Traders can start from all the stocks in the market for their investment analysis. More frequently, s/he focuses on a few sectors (per funder's requirements) and picked several stocks from each sector to construct a balanced initial stock portfolio. Our system provides an interface to allow users to manually determine the initial stock portfolio from the list of stocks organized by sectors. Initial factor selections are automatically recommended based on the stock portfolio. (\textbf{R1.1}) 

\textbf{T2. Evaluation of factors through visualization.} Our system illustrates the performance of individual factors in a comprehensive visualization interface. Users can visually identify valuable factors by examining key dimensions such as stability and sensitivity. The stock returns of quantitative models are also displayed. (\textbf{R1.2})

\textbf{T3. Interactive joint factor refinement and portfolio adjustment.} Traders select factors and stocks to construct investment portfolios in an interactive and iterative manner. Upon each refinement of factors or stock portfolio, the stock return of the resulting model is displayed immediately, which provides visual evidence for user's heuristic factor search. (\textbf{R2})

\textbf{T4. Evaluation and comparison of portfolio returns.} To compute portfolio returns, quantitative models are automatically built up over selected factors (\textbf{R3.1}). Traders compare the temporal dynamics of returns from alternative portfolios and select the best portfolio in a standard backtesting (\textbf{R3.2}).

\bsubsec{System Design}{Pipeline}

\begin{figure}[t]
\centering
\includegraphics[width=0.9\linewidth]{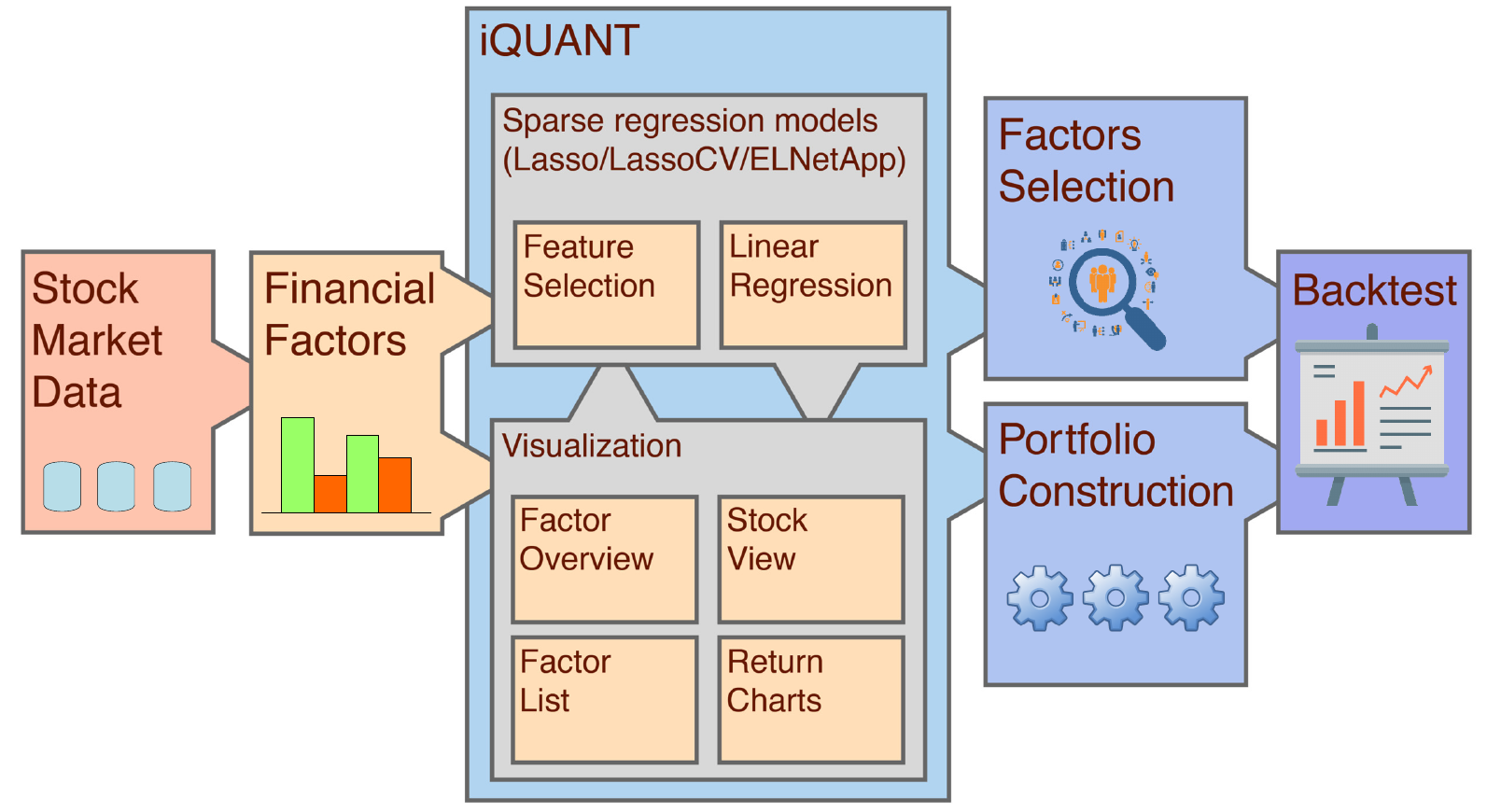}
\vspace{-0.13 in}
\captionsetup{labelfont=bf,textfont=it}
\caption{\codename system pipeline composed of two modules: the sparse regression model for feature selection and prediction, and the visualization of factors and stocks for factor selection and portfolio construction.}
\vspace{-0.15 in}
\label{fig:system_overview}
\end{figure}

We propose \codenamenospace, a visual analytics system supporint all the four user tasks defined above. \codename features a pipelined design as shown in \rfig{system_overview}. In the first stage, the system takes three types of stock market data. The first type is stock transaction data, notably the stock daily opening/closing prices. The second type is market information, including but not limited to financial statements (total assets, debts, cash flows, etc.), market news, and financial reports. The third and most critical part of data are factors computed from stock transactions and market information, which is used in our system for user selection and model building. In this application, we apply data from a major stock market consisting of 3,000 stocks during 30 years (1990-2018). The data is collected from both external sources (e.g., open APIs for transaction data) or by internal computations (e.g., financial factors). 
The underlying technique of our system can be easily extended to support other stock markets.

In the second stage of \rfig{system_overview}, we integrate a sparse regression model to carry out factor selection and return prediction simultaneously. The performance of selected factors and predictive models are visualized in \codename. The result of interactive factor investing is evaluated in a standard backtesting view in the final stage.

The \codename system is a web-based full-stack application. All data are stored in MongoDB. The front-ends use Vue.js and D3.js libraries. 
The sparse regression model is time-consuming. However, it only takes $\sim$20 seconds in the back-end to compute a sector (dozen) of stocks and respond to the web browser. All performance is measured in a laptop requesting a four-CPU cloud server. We expect to upgrade the server to further reduce computation time.

\vspace{-0.05 in}
\bsec{Predictive Analytics}{Analysis}

\bsubsec{Sparse Regression Model}{Model}


The interactive factor investing approach takes two steps to make final decisions on portfolio adjustment. First, the collection of financial factors are used to build models on the pool of candidate stocks for investment. Second, traders evaluate the predicted stock returns by the model and decide on the portfolio adjustment. In \codenamenospace, the second step is achieved by user interactions with the interface to understand the model output, and using existing trading strategies/algorithms (beyond the scope of this paper). In this part, we describe a suite of sparse regression models applied in the first step to jointly recommend factors (i.e. feature selection) and predict stock returns (i.e., model optimization). \rebuttal{The sparse regression model is a popularly-used one in industry. Users can replace it with other statistics/machine learning models with similar output format. We include the sparse regression model to demonstrate the workflow of \codenamenospace.}

Consider the model for a stock $\Upsilon$ during a time period of $\Gamma = [T+1, T+L]$, where $L$ is the length of $\Gamma$ and any time $t \in \Gamma$ indicates the $t$th trading day. $\Gamma$ is further partitioned into $N$ trading cycles: $\{[T+1,T+D], \cdots, [T+(N-1)D+1, T+ND]\}$, where $D$ indicates the length of a trading cycle by days, $D = L/N$. Take the first trading cycle $\tau_1=[T+1,T+D]$ as an example, the market information in a previous training period $[1, T]$ is used in the predictive modeling. $T$ denotes the fixed length of training period by days. Similarly, for the modeling of the $i$th trading cycle $\tau_i=[T+(i-1)D+1,T+iD]$, the training period of $[(i-1)D+1, (i-1)D+T]$ is used.

On stock $\Upsilon$, denote the actual daily stock returns by percentage in the time period $[1, T+L+1]$ by $\mathbf{y}=(y_1, \cdots, y_{T+L+1})$. Let the model take $F$ factors as input. Their normalized factor values in $[1, T+L+1]$ compose a matrix $\mathbf{X} = (\mathbf{x_1}, \cdots, \mathbf{x_{T+L+1}})$ where $\mathbf{x_i} = (x_{i1}, \cdots, x_{iF})'$. The modeling of stock $\Upsilon$ in $\Gamma$ takes $N$ steps, corresponding to the trading cycles of $\tau_1, \cdots, \tau_N$, respectively. Consider the $i$th trading cycle $\tau_i$, a sub-vector of actual stock returns is used as outcomes, denoted by $\mathbf{y^{(i)}}=(y_{(i-1)D+2}, \cdots, y_{(i-1)D+T+1})$. A sub-factor-matrix is used as the design matrix (predictors), denoted by $\mathbf{X^{(i)}} = (\mathbf{x_{(i-1)D+1}}, \cdots, \mathbf{x_{(i-1)D+T}})$. We fit a sparse regression model:
\beq{LinearModel}
\mathbf{y^{(i)}} = \mathbf{w^{(i)}}'\mathbf{X^{(i)}}+b^{(i)} + \mathbf{\epsilon^{(i)}}
\eeq
Subject to the objective function: 
\beq{NLL-Lasso}
Minimize~~Lasso\_NLL = NLL + \frac{1}{2}\lambda[\alpha ||\mathbf{w^{(n)}}||_1 + (1-\alpha) ||\mathbf{w^{(n)}}||_2^2]
\eeq
where $\mathbf{w^{(i)}}=(w_{1}^{(i)}, \cdots, w_{F}^{(i)})'$ denotes the weight vector of financial factors in the fitted model, $b^{(i)}$ is the bias, $\mathbf{\epsilon^{(i)}}$ is the error term, $||\cdot||_1$ ($||\cdot||_2$) is $L_1$ ($L_2$) norm of the vector. $D$ is set to 21 (average number of monthly trading days). The best weight vector is computed by minimizing the negative log likelihood (NLL) of a linear regression term plus $L_1$ and $L_2$ penalty terms of the weight vector. The penalty terms shrink the weight of unimportant factors to zero. On each trading cycle, the model is applied to predict daily stock returns.

\begin{figure*}[ht!]
\centering
\input{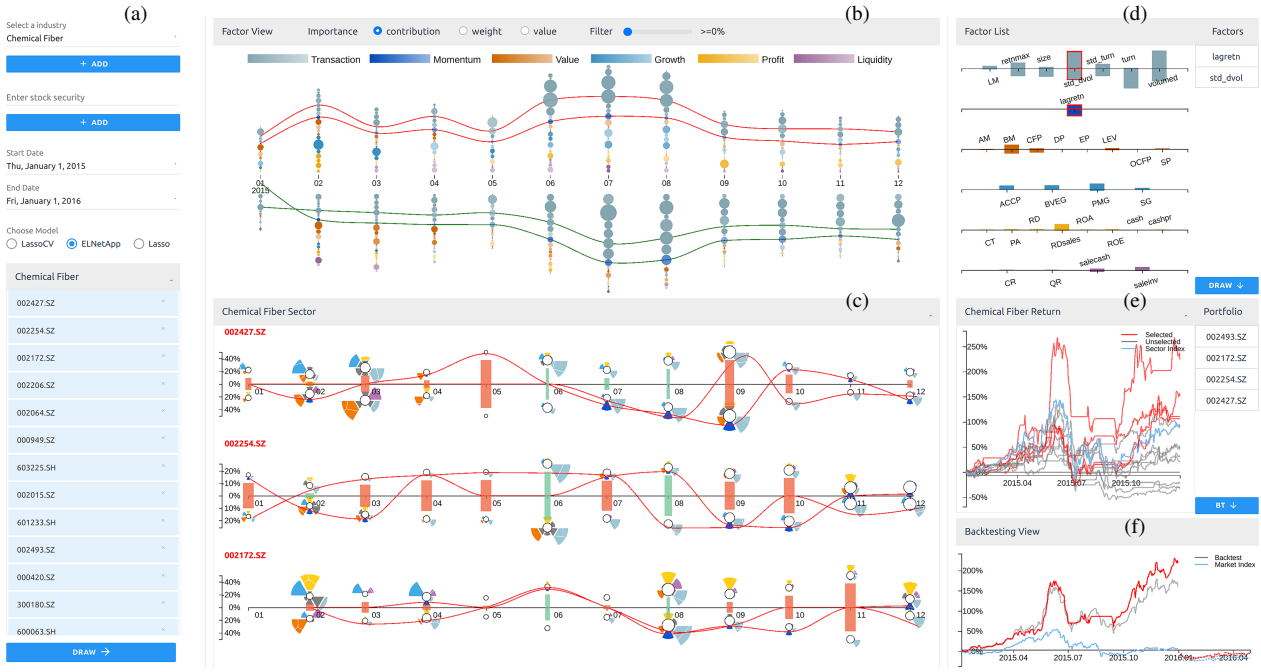}
\captionof{figure}{Interactive selection of four stocks from the Chemical sector in 2015 using a two-factor model (\textbf{lagretn}$+$\textbf{std\_dvol}): (a) control panel; (b) factor view; (c) stock view by sectors; (d) factor list for selection; (e) stock return view for portfolio construction; (f) backtesting view for evaluation of returns. 
\vspace{-0.2 in}
\label{fig:teaser}}
\end{figure*}

Note that $\lambda$ and $\alpha$ in \req{NLL-Lasso} control the degree of model sparsity and weights between $L_1$ and $L_2$ terms. In this work, we provide three choices of modeling: $\alpha=1$ (without the $L_2$ term), the famous \textit{Lasso} model \cite{Lasso} with a fixed $\lambda$; \textit{LassoCV} model using 10-fold cross-validation to determine $\lambda$ of the Lasso model \cite{LassoCrossValidation}; and $\alpha=0.5$, a balanced Elastic Net model (\textit{ELNetApp}) \cite{ElasticNet}. The trade-off of the three models in predictive performance and the number of selected features can be found in recent literature \cite{Lasso}\cite{LassoCrossValidation}\cite{ElasticNet}.

The sparse regression model achieves feature selection ($L_1$ and $L_2$ penalty terms) and model optimization (regression term) simultaneously. This is extremely helpful to alleviate the overfitting effect in the factor investing case when the number of predictors ($F=56$ in our implementation) is comparable to the number of training samples ($T=200$). Though we apply a basic linear regression for stock prediction, the sparse models with $L_1$ and $L_2$ penalties can be easily extended to support more sophisticated models such as nonlinear models using customized kernel functions.

\vspace{-0.15 in}
\bsubsec{Financial Factors}{Factor}
\label{sec_factor_importance}

For factor computation, we adopt the result of a latest white paper on the target stock market \cite{TsinghuaSurvey}, which recommends six types of 56 factors
\eurovis{with each factor characterizes a key aspect of a stock:}

\begin{itemize}
\item \textbf{Transaction friction}: 17 factors delineating overall status of a company (e.g., firm size by its market value, age of a firm) and transaction statistics (e.g., volatility, risks measured by stock returns compared with the market return);
\item \textbf{Momentum}: 5 factors computing daily stock returns in the recent 6/12 months, the change of momentum, and specially defined momentums;
\item \textbf{Value}: 8 factors including the famous book-to-market ratio proposed by Fama and French \cite{ThreeFactor}, the asset-to-market ratio, and other company performance related ratios;
\item \textbf{Growth}: 11 factors describing the growth of a company's asset, debt, market value, sales, profit, tax, etc;
\item \textbf{Profitability}: 8 factors quantifying return on equity, return on asset, and other factors related to company profits;
\item \textbf{Liquidity}: 7 factors characterizing the liquidity performance of a company, including current ratio, quick ratio, cash flow to debt ratio, etc.
\end{itemize}


For the above factors, \eurovis{the ranges of their values differ significantly from each other. Also, their values can change dramatically along with the stock market. \rebuttal{Since all the factor values are normalized initially, the relative importance of each factor can be assessed, }
which can facilitate the factor analysis}. The sparse regression model in \codename outputs three measures to delineate the importance of each factor in the prediction. Take the $j$th factor in the $i$th trading cycle \eurovis{(e.g., daily, monthly or quarterly)} as an example, \eurovis{the three measures are as follows:}

\begin{itemize}
\item Factor \textbf{weight} learned in the model, denoted by $w_{j}^{(i)}$;
\item Factor \textbf{value} changing in a daily basis, denoted by $\mathbf{X^{(i)}}$;
\item \textbf{Contribution} of factors to the prediction of daily stock returns, denoted by $\sum_{k=T+(i-1)D+1}^{T+iD}w_{j}^{(i)}x_{kj}$.
\end{itemize}

By collecting the feedback from traders in our pilot study,
\eurovis{we further extracted three metrics that are widely-used in industry to delineate the utilization of each factor.}
These two metrics are visualized in \codename to aid the interactive factor selection and stock portfolio construction process. 

\begin{itemize}
\item \textbf{Sensitivity}: a metric that indicates whether the utilization of a factor can be replaced by other factors. The sensitivity is computed by $(\xi_{-j}^{i} - \xi^{i})^+$ where $\xi_{-j}^{i}$ denotes the prediction error by removing the $j$th factor from sparse regression models, $\xi^{i}$ denotes the prediction error of the original model;
\item \textbf{Stability}: a good factor for the predictive model should have stable contributions in a long period of time before confidently applied. Quantitatively, it is measured by the number of times the factor's contribution to a stock's prediction flips from negative to positive or from positive to negative.
\end{itemize}
\vspace{-0.05 in}
\bsec{Visualization}{Vis}



\bsubsec{Control Panel}{sec-control-panel}

The \textit{control panel} in Figure~\ref{fig:teaser}(a) allows users to select an initial pool of stocks for analysis (\textbf{\textit{T1}}). S/he can input the stock code to select an individual stock or choose an entire sector of stocks from a sector list. All the selected stocks will appear in the bottom part of the control panel, aggregated by sectors. Next, users need to specify a time period and a model for factor investing. Finally, users click the ``Draw'' button to visualize factors related to the selected stocks.

\vspace{-0.1 in}
\bsubsec{Factor View}{sec-factor-overview}
%

%

\textit{Factor view} (Figure~\ref{fig:teaser}(b)) is designed to provide traders with a quick overview of financial factors relevant to the selected stocks and time period. These factors are initially recommended by sparse models in \codename (\textbf{\textit{T1}}). The model measures the importance of each factor by its weight, value, or contribution to the prediction of stock returns. By default, traders first look at the \emph{contribution} measure which covers both its current value and the overall weight. The contribution can be either \emph{positive or negative} depending on the relationship between the factor and stock returns. According to our expert study, they also consider several other metrics in the factor evaluation and selection process, including the temporal \emph{stability} of factors and their \emph{sensitivity} in the model. (\textbf{\textit{T2}})

To meet the above design requirements, we propose a customized visual design to integrate time-varying factor importance metrics in a single factor view. As shown in Figure~\ref{fig:teaser}(b), the x-axis of the view indicates a timeline divided into multiple time units (months by default). Each factor is visualized as a series of filled circles in these time units. On each circle, the color hue encodes factor type, the color opacity encodes the factor sensitivity to the model, and the circle radius represents the aggregated factor importance to all the selected stocks. Three importance metrics can be displayed by the checkboxes in the top row of the factor view: weight, value, and contribution (default). All the circles on the same time unit are vertically aligned by a fixed order of their corresponding factors to facilitate factor tracking and comparison along the timeline. As each factor can have a positive or negative contribution in different time units, we depict positive and negative factors as circles above and below the x-axis respectively.

User interactions are designed in the factor view to help identify useful factors for modeling. When a user hovers a factor circle, all the circles representing the same factor are linked by a curved line to reveal its stability over time.  (Figure~\ref{fig:teaser}(b)). \rebuttal{Also, 
a tooltip will be displayed, showing the detailed information of the factor circle such as factor type, name, sensitivity and importance metric.}
The user can interactively filter factors to reduce visual complexity.



\begin{figure}[t]
\centering 
\includegraphics[width=0.95\linewidth]{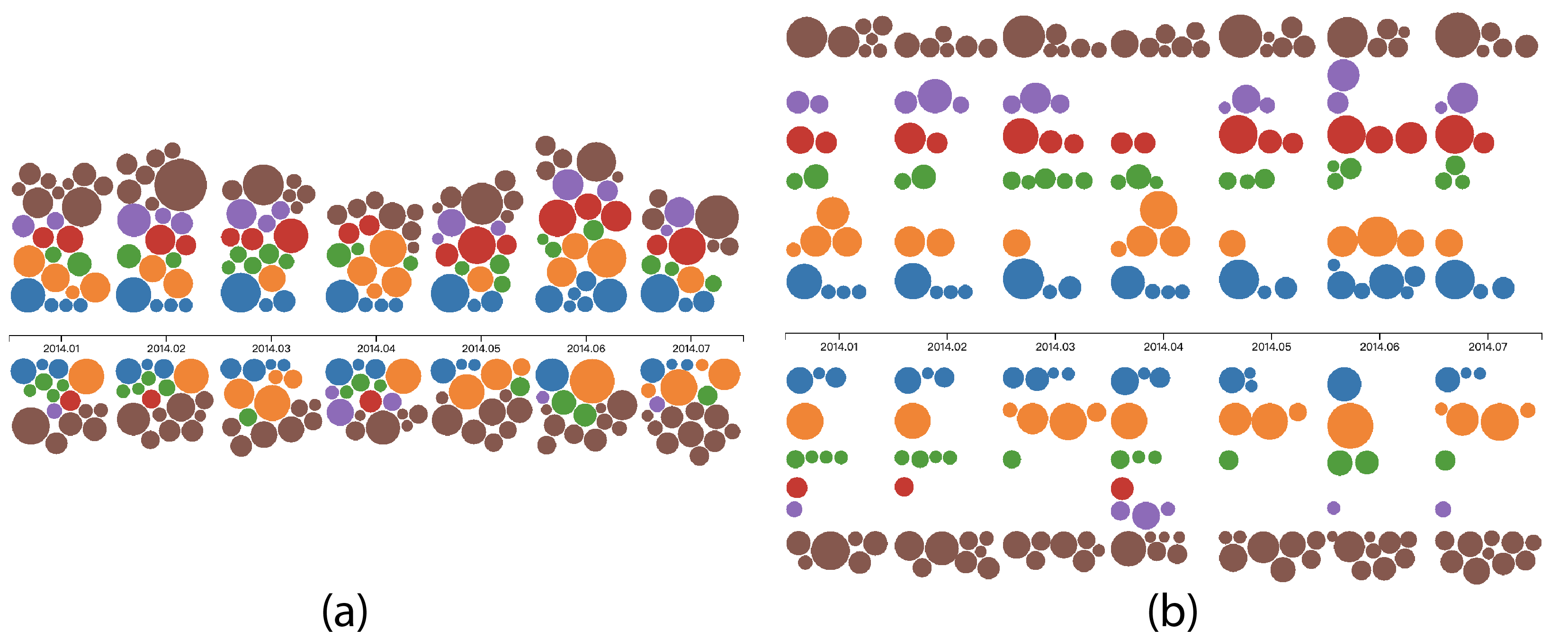}
\vspace{-0.15 in}
\caption{Alternative designs of the circle layout in the factor view: (a) circle packing; (b) layered circle packing.}
\vspace{-0.23 in}
\label{fig_factor_alternatives}
\end{figure}

\textbf{Alternative designs} of the factor view have been considered. The line chart is the most intuitive form to visualize factor importance over time. When we have as many as 56 factors, line charts suffer from severe visual clutter. Stacked flow chart~\cite{byron:2008:stackedflow} is another possible visual design. A quick overview of factor composition can be provided, but it is not good at displaying the temporal dynamics of individual factors (especially for unimportant factors). For the circle layout, we also experimented with circle packing in which positive and negative factors are placed separately. The circle packing leads to more compact visualization (Figure~\ref{fig_factor_alternatives}(a)). 
However, our domain experts complained that the design made it difficult to track individual factors across the timeline. 
The relative position of the same factor circle varied a lot across different time units.
In an improved layered circle packing (Figure~\ref{fig_factor_alternatives}(b)), \rebuttal{which is similar to a matrix design,} the same type of factors is aligned horizontally. It facilitates tracking of the same factor, but the space usage is inefficient and the connecting lines on factors lead to heavy crossings. \rebuttal{The current design is a good trade-off between space usage and effectiveness compared with other alternative designs.}

\vspace{-0.1 in}
\bsubsec{Stock View}{sec-stock-view}
\textit{Stock view} (Figure~\ref{fig:teaser}(c)) splits the aggregated factor time series in the factor view according to their importance to individual stocks. This helps to evaluate the factor performance among different stocks and sectors (\textbf{\textit{T2}}). The visual design is challenging, as the space for each stock is limited and there could be many factors relevant to a stock. On each factor, there are also multiple attributes to be visualized, e.g., the magnitude and polarity of importance (contribution), the model bias, and the trend of the stock price.



In our design, the visualization of each stock follows a similar layout to the display in the factor view. As shown in Figure~\ref{fig:teaser}(c), the x-axis is a timeline divided into units. At each time unit, a stock bar is drawn (\rebuttal{the yellow bar of Figure~\ref{fig:ChemSectorPortfolio1}} (d)), whose height represents the ratio of stock price change in this time unit and width represents the rate of prediction error on the price change by models. A red (green) bar indicates a rise (drop) of the stock price. On top of and below each stock bar, there are two \textit{factor glyphs} which summarize the factors having a positive and negative contribution to the price change in the predictive model.

The factor glyph \rebuttal{
at the top and bottom in Figure~\ref{fig:ChemSectorPortfolio1} (d)
} has a circular shape, and is composed of one inner circle and six angular sectors in the outside. The radius of the inner circle encodes the model bias. Each outer sector represents one type of factors, with the size of each sector indicating the sum of all factor's importance in this type. On each factor glyph, the top-5 factors with the highest importance in all factor types are displayed. When more than one \rebuttal{factor} from the same type belongs to the top-5, a Voronoi diagram is drawn to pack these factors in the same angular sector. Each region in the Voronoi diagram corresponds to one factor, with the region size indicating its importance. The sum of all the other factor's importance out of the top-5 category is represented by the grey region close to the glyph center. When users hover a Voronoi region of one factor, a line will be drawn to connect all the Voronoi regions of this factor over time. \rebuttal{Meanwhile, a tooltip showing the factor type, name and importance will also appear}. When the factor does not belong to the top-5 category, the line will be linked to zero position on the x-axis.

\textbf{Alternative designs} such as line chart and streamgraph~\cite{havre2000themeriver} are commonly used for visualizing temporal evolution of variables. In our scenario, the top-5 factors can be quite different in separate time units, making line charts or streamgraph discontinuous over time and hard to track by users.
Commodity designs such as stacked bar charts and pie charts are also considered. Nevertheless, the stacked bar chart is not scalable to support dozens of factors in the same view. The pie chart can clearly show the percentage of each factor in importance, but again the relative position of the same factor can vary a lot in different units, making it hard to compare the evolution of the same factor.
\rebuttal{
\codename is designed for financial factor selection, which makes it inevitable to consider much domain knowledge and incorporate different factors in our design.
This can increase the design complexity and it takes some time for 
user to learn, but most of our users can be well prepared after a training session of about 10 minutes as indicated in \rsec{expertfeedback}, which we would argue is a reasonable learning effort.}





\vspace{-0.15 in}
\bsubsec{Factor List}{sec-factor-list}

\textit{Factor list} (Figure~\ref{fig:teaser}(d)) aggregates factor data together to depict the overall factor performance (\textbf{\textit{T2}}). The six types of factors are listed in rows, with factors in one type uniformly distributed as factor bars in the same row. Each factor bar represents the positive and negative contribution of the factor in the entire time period, using the height of the bar above and below the x-axis respectively. Users can manually select factors by clicking factor bars, then the stock prediction models are re-built using the selected factors after triggering the ``Draw'' button. The factor performance in the stock view is also updated for iterative factor/stock selection. (\textbf{\textit{T3}})

\vspace{-0.15 in}
\bsubsec{Stock Return View}{sec-stock-return}

\textit{Stock return} view (Figure~\ref{fig:teaser}(e)) is organized by sectors. In each expandable sector box, the investment returns of all stocks in that sector are calculated by backtesting and displayed in multiple grey line charts. An additional blue line indicates the market return. Users could select promising stocks in the stock view and examine their returns by the lines highlighted in red. The stocks could be selected in the return view and added into a \textit{portfolio} table. A ``Backtest'' button will trigger the backtesting computation by the selected factors (factor list) and selected stocks (portfolio table).

\vspace{-0.1 in}
\bsubsec{Backtesting View}{sec-back-test}

\textit{Backtesting view} (Figure~\ref{fig:teaser}(f)) evaluates the investment performance of the selected portfolio (\textbf{\textit{T4}}). Backtesting calculates the portfolio returns using historical stock data. The blue line indicates the average return in the market and the grey one indicates the portfolio return. A dotted line after the selected time period represents the predicted return in future. Comparison among multiple portfolios is enabled by adding more portfolio from the stock return view.

\vspace{-0.1 in}
\bsubsec{Multiple Views Interaction}{sec-multiple-views}



{\codenamenospace} supports interactions among multiple views for comprehensive evaluation of factor performance (\textbf{\textit{T2}}) and interactive selection of stocks and factors (\textbf{\textit{T3}}). 

\textbf{Selection of factors}: Users can select/de-select factors by clicking factor circles in the factor view, Voronoi regions in the stock view, and factor bars in the factor list. Each change to the factor selection will trigger updates in all the views simultaneously. 

\textbf{Selection of stocks}: An initial collection of stocks can be selected in the control panel. Users can further compose an investment portfolio by clicking the stock name in the stock view or the return curve in the stock return view. This helps to interactively and iteratively refine stock selection according to their returns or the performance of relevant factors. 

\vspace{-0.15 in}
\bsec{User Experiment}{User}



\bsubsec{Design}{ExpDesign}

We conducted a formal user study to compare the effectiveness of \codename visualization design with a baseline factor visualization using commodity multiple line charts (See the \eurovis{supplementary material} for more details of the baseline factor visualization used). \eurovis{Since there is no efficient tool that can fulfill the tasks mentioned in \textit{Task Characterization}, we chose to compare \codename with the widely-used line charts in the stock market.} In both designs, the time series of financial factors and stock prices were displayed. Users were required to complete multiple tasks involving the selection of factors and stocks for investment, which corresponded well with \textbf{T1$\sim$T3} in \textit{Task Characterization}. In total, 14 expert quantitative traders were recruited as subjects, who were paid for their valuable time. We applied a between-subject design that subjects were randomly assigned to two groups of equal size. The seven subjects in each group will complete all the tasks with only one visualization, \codename or baseline. The experiment was composed of two sessions. In the first training session, subjects were presented with a tutorial and allowed a trial usage of \codename to be tested. In the subsequent formal test session, each subject took seven tasks. To alleviate the influence of financial data on the result and increase statistical power, we presented two stock data. Each subject was required to complete all the tasks with both data using the tested visualization. The full documents of the experiment design are provided in Appendix A. The user tasks can be categorized into four groups.

\emph{Q1/Q2 (\textbf{factor selection by contribution}): Among all the factors, which factors have the highest positive/negative contribution to the return of a particular stock during a given time period?}

\emph{Q3 (\textbf{factor selection by stability}): Among all the factors, which factors have the highest unstable contribution to the return of a particular stock during a given time period?}

\emph{Q4/Q5 (\textbf{factor selection by contribution to a sector of stocks}): Among all the factors, which factors have the highest positive/negative contribution to the return of all stocks in a particular sector during a given time period?}

\emph{Q6/Q7 (\textbf{stock selection by factor contribution}): Among all the stocks in a given sector, whose return receive the highest positive/negative contribution from a selected factor during a given time period?}

For each question, the subject was required to provide $1\sim2$ answers. Because of the between-subject design, subjects were not asked for their subjective rating on each visualization (no within-subject comparison). Instead, we collected their verbal feedback on the pros and cons of each design.


\vspace{-0.12 in}
\bsubsec{Result and Analysis}{ExpResult}

\begin{figure}[ht!]
\centering
\includegraphics[width=0.95\linewidth]{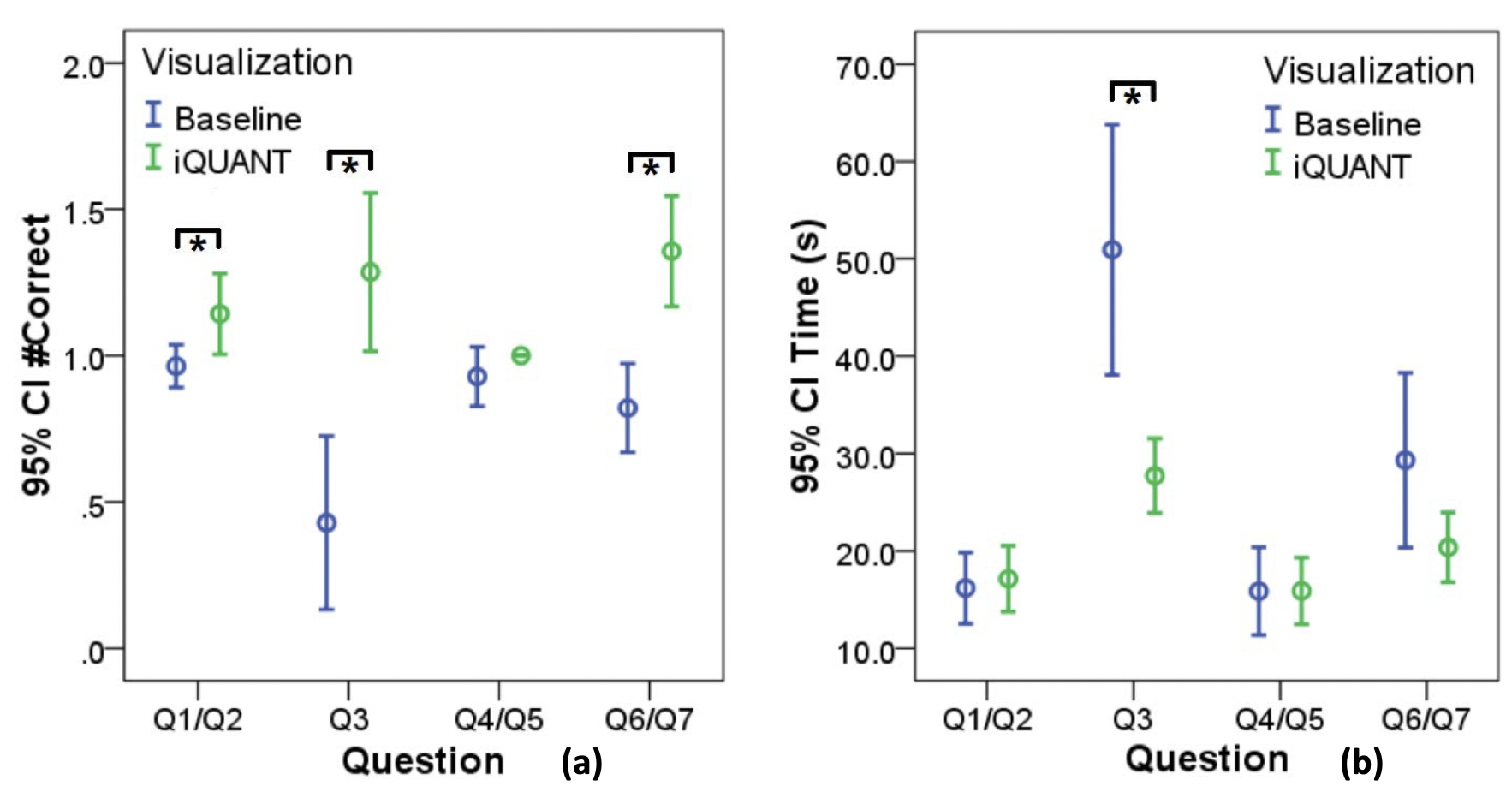}
\caption{User experiment results: (a) the number of correct answers provided by users in two visualization designs; (b) the task completion time.}
\vspace{-0.2 in}
\label{fig:ExperimentResult}
\end{figure}

We compute the accuracy of each user's response to a question by comparing them with top-2 answers. The overall accuracy is measured by the number of correct answers to each question. In \rfig{ExperimentResult}, we depict the accuracy and completion time of four groups of seven questions. Because both accuracy and time do not follow normal distributions, we apply the Mann-Whitney test to analyze.

On task accuracy (\rfig{ExperimentResult}(a)), the result reveals that, for factor selection by contribution (Q1/Q2), by stability (Q3), and stock selection by factor contribution (Q6/Q7), the accuracy of \codename is significantly higher than the baseline design: $U = 324.0,~p = .024$ for Q1/Q2; $U = 30.0,~p = .001$ for Q3; $U = 207.0,~p < .001$ for Q6/Q7. The average task accuracies are 1.14$\pm$0.14 (\codenamenospace) and 0.96$\pm$0.07 (baseline) for Q1/Q2; 1.29$\pm$0.27 (\codenamenospace) and 0.43$\pm$0.3 (baseline) for Q3; 1.36$\pm$0.19 (\codenamenospace) and 0.82$\pm$0.15 (baseline) for Q6/Q7. For factor selection in a sector (Q4/Q5), the accuracy difference between the \codename and baseline is not significant ($U =364.0,~p = .15$), though the average accuracy of \codename (1.0$\pm$0.0) is higher than the baseline (0.93$\pm$0.1).

On task completion time (\rfig{ExperimentResult}(b)), the result reveals that, for factor selection by stability (Q3), the completion time using \codename is significantly shorter than the baseline design ($U = 30.5,~p = .002$). The average completion times are 27.2s$\pm$3.8s (\codenamenospace) and 50.9s$\pm$12.8s (baseline). For factor selection by contribution (Q1/Q2), in a sector of stocks (Q4/Q5), and stock selection by factor contribution (Q6/Q7), the completion time using \codename is not significantly shorter than the baseline design: $U = 359.0,~p = .58$ for Q1/Q2; $U = 367.0,~p = .68$ for Q4/Q5; $U = 342.0,~p=.41$ for Q6/Q7. The average completion times are 17.1s$\pm$3.4s (\codenamenospace) and 16.2$\pm$3.6s (baseline) for Q1/Q2; 15.9s$\pm$3.4s (\codenamenospace) and 15.9s$\pm$4.5s (baseline) for Q4/Q5; 20.4s$\pm$3.6s (\codenamenospace) and 29.3s$\pm$9.0s (baseline) for Q6/Q7.

The study result indicates that, on most tasks, \codename leads to higher task accuracy than the baseline design. The only exception happens on the factor selection task for a sector of stocks (Q4/Q5). It is found that in the stock data used for these two questions, the factor having the highest positive/negative contribution is quite separated from all the other factors. Hence, only the top-1 factor is used as the correct answer, and in both designs, subjects could easily identify this factor (a 100\% accuracy for \codenamenospace). For similarly easy tasks (Q1/Q2 and Q6/Q7), though the baseline design in most case helps to identify one (but only one) top factor, the accuracy of \codename is significantly higher because subjects could quickly discover more top factors in comparison to the baseline design. For the hard task of selecting factors by stability (Q3), 
the advantage of \codename in accuracy becomes the highest. Meanwhile, on completion time, \codename is favored only in the hard task of Q3, for which the \codename design can better support temporal comparison and analytics of factor contribution. For the other tasks, the completion time is similar in both \codename and baseline design.

\begin{figure*}[ht!]
\centering
\input{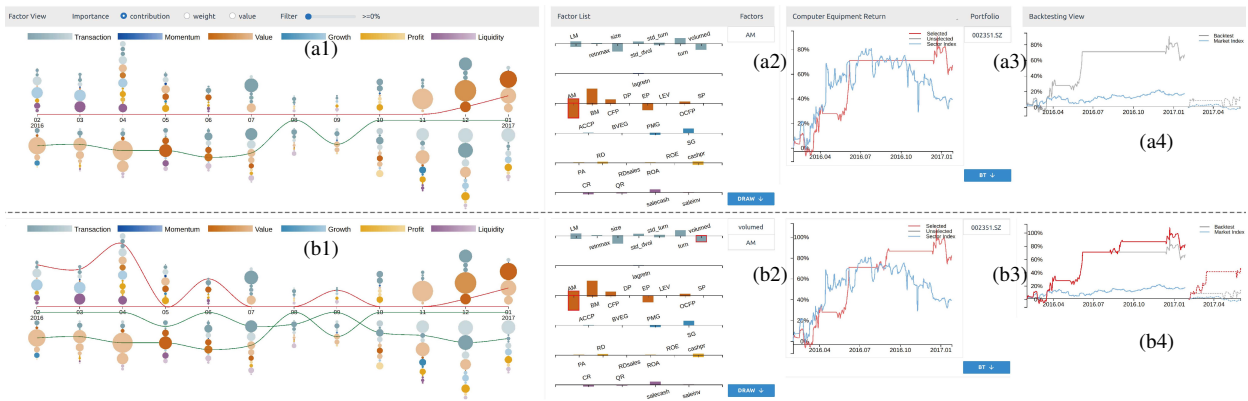}
\vspace{-0.15 in}
\caption{The selection of investment time during 2016.2-2017.1 on 002351.SZ by: (a) a single-factor model (\textbf{AM}); (b) a two-factor model (\textbf{AM}+\textbf{volumed}). \eurovis{
The manager switched from a single-factor model to a two-factor model to optimize his investment strategy.
}}
\vspace{-0.15 in}
\label{fig:Case_selection_time}
\end{figure*}

\vspace{-0.13 in}
\subsection{Expert Feedback}\label{sec:expertfeedback}

In the user experiment, experts also provided valuable feedback about several aspects of \codename.


On positive aspects, some experts stressed that \codename presented an effective way to visually compare salient factors among different sectors of the market. In practice, the factor constituent and dynamics are truly different across sectors, which is of great importance to factor investing. As supported by the study result, experts also found \codename to be useful in analyzing the temporal dimension of factors, as well as the relationship between factors and stocks, which are exactly our design goal. Overall, subjects reported the \codename interface to be user-friendly, as our design was similar to their familiar and popular tools. Some experts even commented that our design could be part of standard stock market analysis tools (e.g., the work from Bloomberg \cite{BloombergVis}).


On the other hand, it was suggested that the interface could incorporate more statistical information according to the industry standard (e.g., the change of stock holdings in the backtesting view). The Voronoi map for top-5 factors could be reduced to only show the top-3 or less, in order to better reveal the most salient signal for the stock market. More detailed visual encoding could be applied to distinguish the factors of the same type, which are now depicted in the same color. Notably, we thought the complexity of the interface could raise learning issues. During the study, most subjects could be well prepared after a 10-minute training session, as indicated by the high task accuracy in applying \codenamenospace. 

\vspace{-0.1 in}
\bsec{Case Study}{Eva}




\bsubsec{Selection of Investment Time}{SelectTime}

We first invited a fund manager to select the best investment time for individual stock with \codenamenospace. He looked at the year of 2016.2-2017.1 when the market index rose by 15.7\%. From a list of candidates, the manager examined each stock until a feasible investment time for excess return is identified.

The manager reported stock in the Computer sector (code: 002351.SZ) for which the composite index of the whole sector rose by 2.5\%. In his analysis, the Lasso model was selected. As shown in Figure~\ref{fig:Case_selection_time}(a1), the initial factor view displays a summary of important factors recommended by the Lasso model. In the factor list view (Figure~\ref{fig:Case_selection_time}(a2)), the manager identified several useful factors having large contribution to the stock return. He hovered these factors one by one to examine their performance in the given period. The \textbf{AM} factor (asset-to-market ratio) was chosen because its contributions are stable in the first six months: all negative to the rise of stock prices. There are two months when the factor sensitivity is high, as indicated by opaque colors. By this one-factor model, the investment return (red curve in Figure~\ref{fig:Case_selection_time}(a3)) surpasses the stock price change (blue curve) by 26.5\% at the end of the period. A maximal return of 11\% can also be forecast for the next three months, as shown by the grey dotted curve in the backtesting view of Figure~\ref{fig:Case_selection_time}(a4). Notably, the manager suggested applying the model from June when the \textbf{AM} factor has been stable for four months and enjoys a high sensitivity in the latest month (May). In case this strategy was adopted, an excess return of 39.1\% could be earned in June when the overall market only rose by 1.6\%.

To optimize the investment for the whole year, the manager explored other factors that can characterize the price change in the second half of the year. He found that the \textbf{volumed} factor was complementary to \textbf{AM}, as the \textbf{volumed} factor contributed to the model mainly from July (Figure~\ref{fig:Case_selection_time}(b1)). By applying a \textbf{AM}+\textbf{volumed} two-factor model, a return of 81.8\% can be obtained by the end of the period, which is 14.5\% more than the one-factor model (Figure~\ref{fig:Case_selection_time}(b3)). Notably, the predicted return for the next three months increases to 40.9\% with the new model (Figure~\ref{fig:Case_selection_time}(b4)).

\begin{figure*}[ht!]
\centering
\input{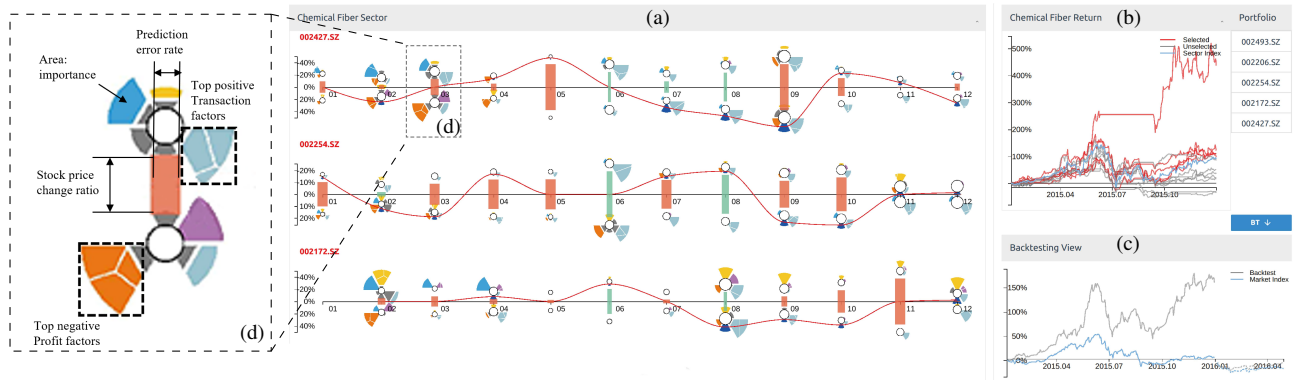}
\vspace{-0.15 in}
\caption{Portfolio construction from the Chemical sector using the single factor (\textbf{lagretn}) in 2015: (a) \textbf{lagretn} factor contributed similarly to theses three stocks; (b) selected high return stocks for portfolio; (c) backtesting results from the portfolio; \rebuttal{(d) a detailed illustration of the Voronoi-based glyph.} }
\vspace{-0.15 in}
\label{fig:ChemSectorPortfolio1}
\end{figure*}

\vspace{-0.15 in}
\bsubsec{Selection of Stock Portfolio}{SelectPortfolio}




In the second scenario, we invited the fund manager to select a portfolio of stocks to invest from an initial stock pool in a given time period. The sector-based division was adopted. The manager analyzed every candidate sector, and combined the selected stocks in each sector into a final portfolio for investment. The year of 2015 was considered when the market experienced a spike with the maximal index climbing as high as 60\% in the first six months. The yearly growth was 9.4\% after the downturn in the second half of the year. It is thus an appropriate scenario to test whether our tool can improve factor investing even in rapidly changing markets.

The manager first studied the Chemical sector which had a similar trend to the market in 2015. The factor view in Figure~\ref{fig:teaser}(b) visualizes an initial collection of factors using the ELNetApp model. It can be seen in the factor view that many factors are quite stable over time. Different from the study on individual stocks, for portfolio construction, the manager would like to select both valuable factors and a set of stocks whose selected factors behave in a similarly stable and effective way. Our tool provides support with an interactive selection process. Users can click on each important factor in the factor list (Figure~\ref{fig:teaser}(d)) and scroll in the stock view (Figure~\ref{fig:teaser}(c)) to identify candidate stocks. By this process, the manager found factor \textbf{lagretn}, which had a moderate overall contribution to stock returns of the sector (Figure~\ref{fig:teaser}(d)) and the contribution was stable on several individual stocks. As shown in  Figure~\ref{fig:ChemSectorPortfolio1}(a), the selected stocks share similar patterns on \textbf{lagretn}. When the stock price started to rise again from September (notice red/green bars in Figure~\ref{fig:teaser}(c)), the contribution of \textbf{lagretn} reversed to negative and stayed for few months, as shown by the connecting lines in Figure~\ref{fig:ChemSectorPortfolio1}(a). It is found in Figure~\ref{fig:ChemSectorPortfolio1}(b) that five stocks (red lines) sharing this pattern on \textbf{lagretn} receive excess returns above the sector average (blue line). The prediction performance in Figure~\ref{fig:ChemSectorPortfolio1}(c) (grey dotted line) also surpasses the market average (blue dotted line).

The manager proceeded to interactively examine all the five selected stocks. In the factor list view, he checked again all the candidate factors and found the factor \textbf{std\_dvol}, which had a favorable pattern as the factor \textbf{lagretn} on all the selected stocks. After applying the two factors together (Figure~\ref{fig:teaser}(d)), the stock returns were re-computed and updated in Figure~\ref{fig:teaser}(e). One stock appeared to receive less return after adding the second factor, while all the other four stocks remained superior to the sector average. In the backtesting view (Figure~\ref{fig:teaser}(f)), the final portfolio of two factors and four stocks (red lines) turned out to be better than the single-factor portfolio (grey lines) in both the backtest performance in 2015 (solid lines) and an outlook in the first three months of 2016 (dotted lines).

In another study, we asked the manager to work in the Household Appliances sector in the same year of 2015. By repeating a similar process, he was able to build a two-factor model of \textbf{size}$+$\textbf{retnmax}. Three stocks having stable contributions were selected. The portfolio returns in both 2015 and 2016 (outlook) are better than the market average. We omit screenshots due to the space limit.

By comparing the visual analysis result in the above two sectors, it is noticed that the Appliance sector is almost exclusively influenced by the transaction factors, while the Chemical sector is influenced by transaction factors and other types of factors such as momentum, value, and growth (Figure~\ref{fig:teaser}(d)). We hypothesize that because of their relationship, the Chemical sector is not only affected by transactions in the stock market but also linked to the price of crude oil, which is independently traded.

\vspace{-0.15 in}
\section{Discussion}





\textbf{The sparse regression model and its flexibility.} In our system, users routinely start from an initial set of factors recommended by the build-in sparse regression model, which potentially limits the flexibility of factor investing by \codename. This issue is alleviated from three directions: in the factor list view (\rfig{teaser}(d)), users can manually select any set of factors as the starting point of analysis; the sparsity parameter of the regression model ($\lambda$) can be tuned in the backend to offer more or less initial factors; finally, the \codename system is fully compatible with any other sparse models or a combination of feature selection and prediction models. 


\textbf{The joint human-model approach and its advantage.} We have mentioned in previous sections, the model-only approach in factor selection suffers from the interpretability issue. Traders need to understand the factor before applied in practice. Also, a human is better in a heuristic search of the best combination within the vast factor space. More importantly, in the financial domain, there is no model that works in the long term. It will be critical to have human stand by models, and would be better to early adjust accordingly.

\textbf{The visual design and its learning curve.} Notice that stock traders may not have a strong background in visualization. Thus, we tried to strike a balance between the intuitiveness and expressiveness of visualization. On one hand, classical designs such as bar charts and line charts are employed in \codename, due to their intuitiveness and users' familiarity. On the other hand, our designs are tailored to the target users. For example in the factor view, green/red are used to indicate the decline/rise of stock prices, according to the convention of our target market. The effectiveness of \codename design is confirmed by the user study result and expert feedback. Most subjects found \codename easy to learn and understand.


\textbf{The user study and its design choices.} We evaluated \codename through a formal user study focusing on the performance of factor/stock selection. The study would have accounted for more stages of \codename by looking at the final stock returns, or the accuracy in predicting. We stayed in the scope of evaluating factors for two reasons. First, the stock return by an interactive factor investing system is determined in complex interactions among factor/stock selection, predictive model, and trading strategy. A higher stock return in the study may or may not be directly correlated with a better factor selection design. Second, there is currently no commercial tool that supports end-to-end interactive factor investing. We compare \codename with a baseline design, which is not natively integrated with prediction models and trading strategies.


\textbf{The deployment of \codename and its extensibility.} The investment industry is highly closed as the winning secret can be too valuable to disclose. This closedness prohibits a more practical approach to directly optimize deployed tools. Instead, we developed a homegrown interactive factor investing pipeline and an end-to-end system. Because of the complexity of quantitative investment, we mainly focus on improving the factor selection in the pipeline. 
\rebuttal{We demonstrated the effectiveness of \codename by using the 20-year real stock market data. However, our system can also be deployed to take in real-time data.}
As a next step, we plan to customize and integrate \codename into the production system of our industrial collaborators, in the hope of making real impacts on the industry.
\vspace{-0.15 in}
\section{Conclusion}
This paper presents \codenamenospace, a visual analytics system that seamlessly integrates algorithmic and user-steered feature selection methods for interactive factor investing. \rebuttal{On} the analytics side, we explore the use of sparse regression models that jointly optimizing factor selection and stock prediction processes. 
On the visualization side, we propose multiple coordinated visual designs to comprehensively illustrate importance metrics of each factor, including its positive/negative contribution to models, the factor stability and sensitivity. On top of the prediction model and visualization design, \codename develops an interactive framework that encloses factor refinement and portfolio construction in the same visual analytics loop. With {\codename}, traders can optimize their investment strategy and evaluate them in industry-standard backtesting. We demonstrate the value of \codename 
through one formal user study 
and two case studies, using a \eurovis{real stock market dataset} recording the stock data
in the past 30 years. 
On the selection of investment time and stock portfolio, our domain experts achieved excess returns from both individual stock and sector-based portfolios.





\vspace{-0.15 in}
\section*{Acknowledgment}
We thank the anonymous reviewers for their valuable comments and Prof. Lei Shi for his constructive suggestions on this project. This work is partially supported by a research collaboration grant from Microsoft Research Asia.

\bibliographystyle{eg-alpha-doi}
\bibliography{reference}



\end{document}